\documentclass[cits]{PoS}

\usepackage[intlimits]{amsmath}
\usepackage{amssymb}
\usepackage{bbm}

\title{Spectroscopy with dynamical Chirally Improved quarks}

\ShortTitle{Spectroscopy with CI quarks}

\author{Christof Gattringer$^a$, C. B. Lang$^a$, Markus Limmer$^a$, Thilo Maurer$^b$,
        \speaker{Daniel Mohler}\hspace{1mm}$^a$ and Andreas Sch\"afer$^b$\\
        \llap{$^a$} Institut f\"ur Physik, FB Theoretische Physik,
        Universit\"at Graz, A-8010 Graz, Austria\\
	\llap{$^b$} Fakult\"at f\"ur Physik, Universit\"at Regensburg, D-93040 Regensburg, Germany\\ 
        E-mail: \email{christof.gattringer@uni-graz.at}, 
        \email{christian.lang@uni-graz.at},
        \email{markus.limmer@uni-graz.at},
        \email{thilo.maurer@physik.uni-regensburg.de},
        \email{daniel.mohler@uni-graz.at},
        \email{andreas.schaefer@physik.uni-regensburg.de}}

\abstract{We present recent results of our dynamical simulations with Chirally Improved fermions and report
on new developments in the determination of excited light-quark meson states using interpolators
constructed by applying covariant derivatives on Jacobi-smeared quark sources within the framework
of the variational method.}

\FullConference{The XXVI International Symposium on Lattice Field Theory\\
		 July 14-19 2008\\
		 Williamsburg, Virginia, USA}

\begin{document}

\section{Introduction}
Simulations with full chiral symmetry using Overlap fermions are still in their
infancy. Currently the computational cost of such simulations prohibits
large-scale simulations of fermions fulfilling the Ginsparg-Wilson (GW) relation
exactly. Therefore, we are working with an approximate algebraic solution to the
GW relation \cite{Gattringer:2000js,Gattringer:2000qu}. These so-called Chirally Improved
(CI) fermions are still computationally demanding as the fermion action includes
several hundred terms ranging up to three links distances.

Simulations with CI fermions in the quenched approximation have demonstrated good chiral and
scaling behavior \cite{Gattringer:2003qx}. Interpolators constructed from
Jacobi-smeared sources \cite{Gusken:1989ad,Best:1997qp} of different width as
well as interpolators containing derivatives have been used to extract the
spectrum of light mesons \cite{Burch:2006dg,Gattringer:2008be} within this
approximation. For the extraction of excited states the variational method
\cite{Michael:1985ne,Luscher:1990ck} has been used, which also enables one to
separate contributions from ghost states \cite{Burch:2005wd}.

Here we present a progress report of ongoing dynamical simulations with CI
fermions. The focus will be on the mass spectrum and we will present preliminary
results for both mesons and baryons. Preliminary results on the simulation and
hadrons masses have previously been presented in
\cite{Lang:2005jz,Frigori:2007wa}. For a discussion of other recent results
for excited states see \cite{Morningstar:2008}.

\section{Details of the simulation}

For our simulations with dynamical quarks we use the L\"uscher-Weisz gauge action
and two mass degenerate CI fermions. We use a a standard HMC algorithm
with mass preconditioning \cite{Hasenbusch:2001ne} and a mixed-precision inverter.
Our action  also incorporates one level of stout smearing. Table {\ref{simtable}}
shows the simulation parameters for our configuration ensembles A-C  (lattice size
$16^3 \times 32$).

\begin{table}[htb]
\begin{center}
\begin{tabular}{cccccc}
\hline
\hline
ensemble&	$\beta_\mathrm{LW}$&	$m_0$& HMC time&	$a$[fm]& $m_\pi$[MeV]\\
\hline
A&	4.70&  -0.050&	591&	0.1507(17) &526(7) \\
B&	4.65&  -0.060& 1108&	0.1500(11) &469(4) \\
C&	4.58&  -0.077& 1046& 	0.1440(11) &318(5) \\
\hline
\hline
\end{tabular}
\caption{Run parameters, lattice spacing and pion masses for ensembles A-C of
lattice size $16^3 \times 32$. The scale has been set using the
Sommer parameter $r_{0,\mathrm{exp}}=0.48$ fm.}
\label{simtable}
\end{center}
\end{table}

\begin{figure}[h!]
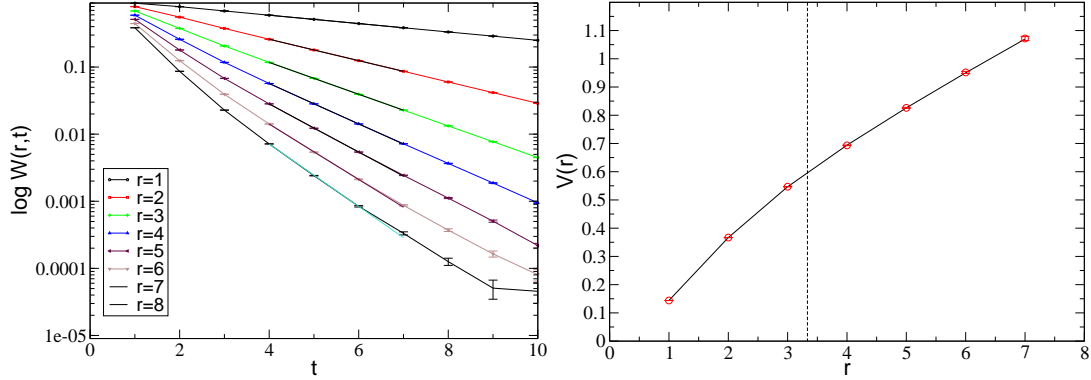

\begin{center}
\includegraphics[height=5cm,clip]{hmc_ci_16x32_b4.58_m-0.077_sum.eps}
\includegraphics[height=5cm,clip]{hmc_ci_16x32_b4.58_m-0.077_pot_v1c.eps}\\
\caption{L.h.s.: Fit to $\ln W(r,t)$ in the range $4\le t \le 7$. R.h.s:  Fit to
the potential in the range $1\le r \le 7$. The data points have been connected
by straight lines to guide the eye.}
\label{potential}
\end{center}
\end{figure}

The lattice spacing has been determined from the static quark potential (for
hypercubic smeared configurations). We fit to the potential
\begin{align}
V_L(r)=A + \frac{B}{r} + \sigma\,r + c_3\, \Delta V(r) \ ,
\end{align}
where $\Delta V(r)\equiv \left[\frac{1}{\mathbf{r}}\right]-\frac{1}{r}$ contains
the perturbative lattice Coulomb potential $\left[\frac{1}{r}\right]$. Using
a value of $r_{0,\mathrm{exp}}=0.48$ fm for the Sommer parameter
we can then determine the lattice spacing as
follows:
\begin{align}
a= r_{0,\mathrm{exp}}\,\sqrt{\frac{\sigma}{1.65+B}}\;.
\end{align}
Figure \ref{potential} shows example fits to the logarithm of the
Wilson loop expectation value $\ln W(r,t)$ and to the potential for ensemble C.

\section{Quark sources for the variational method}

\subsection{Variational method}

For the extraction of hadron masses we use the \emph{variational method}
\cite{Michael:1985ne,Luscher:1990ck}. One uses a matrix of correlators projected
to fixed (here: zero) spatial momentum
\begin{align}
C(t)_{ij}&=\sum_n\big <0|O_i|n\big>\big<n|O_j^\dagger|0 \big > \ ,
\end{align}
and solves the generalized eigenvalue problem
\begin{align}
C(t)\vec{v}_k&=\lambda_k(t)C(t_0)\vec{v}_k\;,\quad
\lambda_k(t)\propto\mathrm{e}^{-tM_k}\left(1+\mathcal{O}\left(\mathrm{e}^{-t\Delta M_k}\right)\right)\ .
\end{align}
At sufficiently large time separation each eigenvalue receives contributions from only a
single mass. At the same time the eigenvectors can serve as a fingerprint to
identify the states when followed over several $t$-values. For a recent discussion 
of the generalized eigenvalue problem in that context see \cite{Blossier:2008tx}.

\subsection{Quark sources}

We construct interpolators from Jacobi-smeared \cite{Gusken:1989ad,Best:1997qp}
quark sources $S$ of different width, with the same choice of the parameters 
$\kappa$ and $N$ as given in \cite{Burch:2006dg}.
\begin{align}
S&=M \; S_0\quad\mbox{with}\quad M=\sum_{n=0}^N\kappa^nH^n\quad\mbox{and} \\
H(\vec{n},\vec{m}\,)&=\sum_{j = 1}^3
\left(U_j\left(\vec{n},0\right) \delta\left(\vec{n} + \hat{j}, \vec{m}\right)+ U_j\left(\vec{n}-\hat{j\,},0\right)^\dagger 
\delta\left(\vec{n} - \hat{j}, \vec{m}\right) \nonumber
\right). 
\end{align}
$S_0$ denotes a point source.
Combinations of such approximately Gaussian sources allow for nodes in the
interpolating operators while using fewer quark propagators than in other
approaches. In addition to these sources we also use derivative quark sources
$W_{d_i}$ by applying covariant derivatives to the Gaussian smeared sources:
\begin{align}
W_{d_i}=D_i\,S\;,\quad
D_i(\vec{x},\vec{y})&=U_i(\vec{x},0)\delta(\vec{x}+\hat{i},\vec{y})-
                   U_i(\vec{x}-\hat{i},0)^\dagger\delta(\vec{x}-\hat{i},\vec{y})\; .
\end{align}

\section{Spectrum results}

In this section we present results from  quenched lattices as well as
preliminary results from the simulation with dynamical CI quarks.

\subsection{Results from quenched lattices}

\begin{figure}[t]
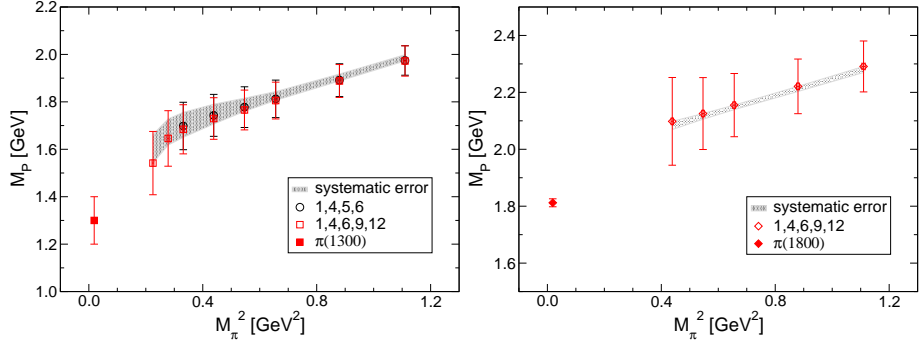

\begin{center}
\includegraphics[width=6cm,clip]{pseudo_first_alternat.eps}
\includegraphics[width=6cm,clip]{pseudo_second_alternat.eps}
\caption{1st and 2nd excitation of $\pi$. The shaded region indicates the
systematic uncertainty due to the choices of fit-range and interpolator
combination.}
\label{quenchedpion}
\end{center}
\end{figure}

Figure \ref{quenchedpion} shows results for the pion channel based on 99
quenched configurations. A clear signal can be obtained for both the first and
the second excitation. The results from the set of interpolators containing
both Gaussian and derivative sources enable us to identify the second excited
state which (on the same dataset) cannot be observed with the Gaussian sources
alone. For more details, some further plots and results from other channels see
\cite{Gattringer:2008be}.

\subsection{Results from dynamical lattices}

For the dynamical CI lattices, the lattice size is $16^3 \times 32$, corresponding to
a spatial extent of $(2.4 \,\textrm{fm})^3\times 4.8 \,\textrm{fm}$. We analyze
every fifth configuration and shift the source positions for consecutive
configurations to reduce autocorrelation. All data presented are of a
preliminary nature. Table \ref{analyzetable} provides an overview of the
current ensembles.

\begin{table}[bp]
\begin{center}
\begin{tabular}{lccccc}
\hline
ensemble& \# conf.s & $a$ [fm]& $m_{\mathrm{AWI}}$ [MeV]& $m_\pi$ [MeV]& $a \,m_\pi\;L$\\
\hline
A&	100/100& 0.1507(17) & 43.0(4)&526(7)&6.4\\
B&	100/200& 0.1500(11) & 35.1(2)&469(4)&5.8\\
C&	100/200& 0.1440(11) & 15.0(4)&318(5)&3.8\\
\hline 
\end{tabular}
\caption{Dynamical CI runs. \# conf.s refers to the number of (independent)
configurations analyzed for mesons or baryons respectively. Values in physical
units have been obtained as described in Section 2.}
\label{analyzetable}
\end{center}
\end{table}

\begin{figure}[tbh!]
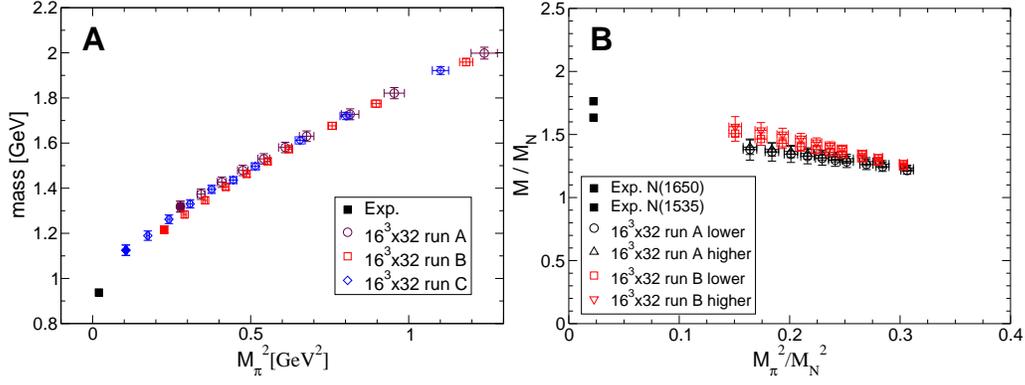

\begin{center}
\includegraphics[height=5cm,clip]{nucleon_masses_draft.eps}
\includegraphics[height=5cm,clip]{nucleon_masses_negative_APE_nucleon_draft.eps}
\caption{A: Mass of the (positive parity) nucleon ground state;
B: Mass of the negative parity ground state and 1st excited
 state.}
\label{nucleon}
\end{center}
\end{figure}

\begin{figure}[tbh!]
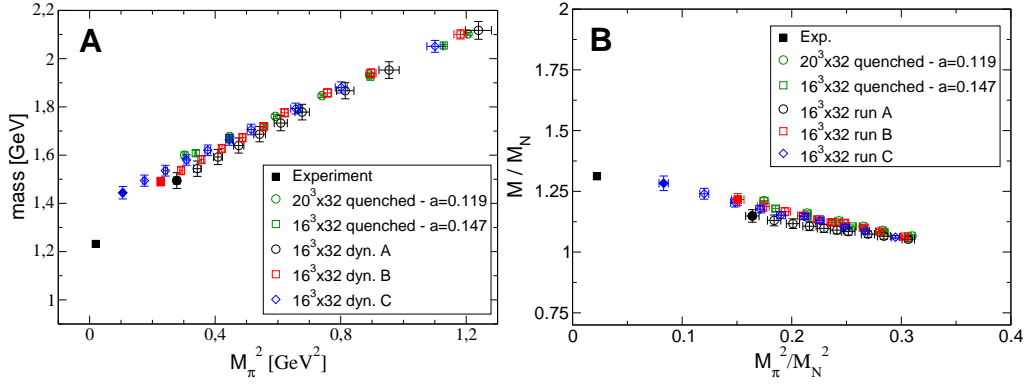

\begin{center}
\includegraphics[height=5cm,clip]{delta_pos_masses_draft.eps}
\includegraphics[height=5cm,clip]{delta_pos_masses_APE_nucleon_draft.eps}
\caption{A: Ground state in the Delta channel in physical 
units; B: Same plotted in units of the nucleon mass.}
\label{delta}
\end{center}
\end{figure}

\subsubsection{Results for baryons}

Figure \ref{nucleon}A  shows results for the mass of the (positive parity) nucleon
ground state. The empty symbols indicate a nucleon made from valence quarks that are
heavier than the sea quarks while the full symbols indicate the fully dynamical,
unitary point for the respective runs. The error bars are purely statistical, also
including the statistical error in the determination of the scale. Systematic errors
from varying fit ranges and combinations of interpolators are not included in the
error bars.  Since we have set the scale individually for each of the parameter sets
this amounts  to a mass dependent renormalization scheme. We therefore cannot use
ChPT extrapolations valid in the  mass independent scheme. More runs at different
masses and volumes will be necessary for reliable extrapolations to the physical
points.

Figure \ref{nucleon}B  shows the ground and first excited state masses observed in the
negative parity channel. While the statistical errors for the different runs show a
deviation a more careful analysis of the systematics due to the choice of
interpolators and fit ranges will be necessary. The quality of the data for run C
was not sufficient to fit these states at the unitary point, therefore it has been omitted from the plot.

Figure \ref{delta} shows the ground state mass in the $\Delta$-channel plotted
in physical units and in units of the nucleon mass. The $\Delta^{++}$ is a
resonance, which complicates the interpretation of the ground state signal. To
get a clear interpretation an analysis of this state in different volumes will
have to be performed.  We observe that when plotting the $\Delta^{++}$ in
units of the nucleon mass the finite volume effects at small pion masses
cancel, thus leading to consistent data for all three runs.

\begin{figure}[tbh!]
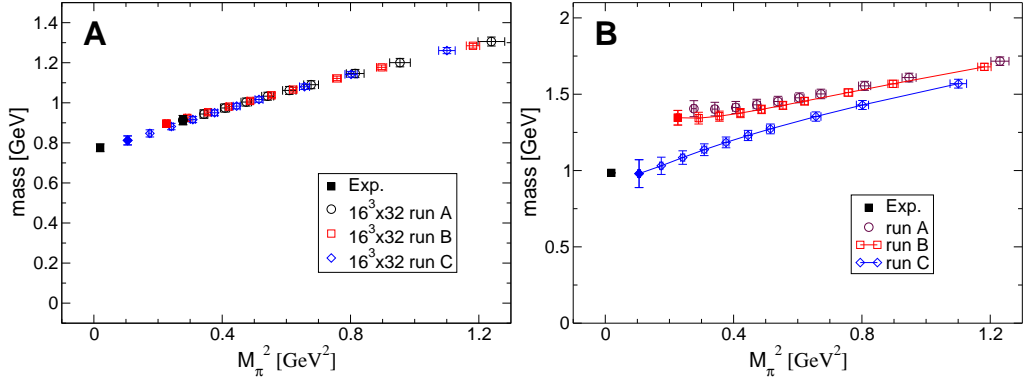

\begin{center}
\includegraphics[height=5cm,clip]{massplot_rho_draft.eps}
\includegraphics[height=5cm,clip]{a0_massplot_00000000110.eps}
\caption{A: Ground state in the $1^{--}$ channel ($\rho$-meson); 
B: Ground state signal in the $0^{++}$ channel for
interpolators 9, 10 in the notation of \cite{Gattringer:2008be}.}
\label{mesonfigure}
\end{center}
\end{figure}

\subsubsection{Results for mesons}

For the mesons the set of interpolators containing derivative sources described
in detail in \cite{Gattringer:2008be} has been used. As can be seen in Table
\ref{analyzetable}, only a subset of configurations was used for this
preliminary analysis. Figure \ref{mesonfigure}A shows some results for the
ground state of the $\rho$ meson. Figure \ref{mesonfigure}B displays results 
for the lowest energy level in the isovector scalar $a_0$ channel for one
possible set of interpolators. We  point out, however, that our data show a
large operator dependent variation. Most quenched results\footnote{Notice that
the analysis of the quenched data is made difficult by artifacts, so called
``ghosts''.} indicate that the ground state in this channel might be the
$a_0(1450)$, in agreement with a tetraquark interpretation of the $a_0(980)$.
Results from dynamical simulations lead to somewhat lower states, possibly due
to the presence of $\pi\eta$ scattering states. A systematic analysis will have
to take into account several volumes to discriminate scattering states from
bound states. One possible interpretation of our results would be a crossing of
the lowest energy level at the quark mass where the $\pi\eta$ state becomes
lower than the $a_0(980)$.

\section{Conclusions}

We presented a progress report on spectroscopy with dynamical CI fermions. We
demonstrated that good signals can be obtained for (most) ground state mesons
and baryons. Emphasis has been put on describing the methods and we are
currently refining these by experimenting with different gauge link smearings.
It would be desirable to perform further runs with the CI action to control both
discretization and finite volume effects and to be able to use a mass
independent renormalization scheme.

\acknowledgments

The calculations have been performed on the SGI Altix 4700 of the
Leibniz-Rechenzentrum Munich and on local clusters at ZID at the University of
Graz. We thank these institutions for providing support. M.L. and D.M. are
supported by ``Fonds zur F\"orderung wissenschaflicher Forschung in \"Osterreich''
(DK W1203-N08). A.S. and T.M. acknowledge support by DFG and BMBF. The work has been
supported by DFG project SFB TR55.

\providecommand{\href}[2]{#2}\begingroup\raggedright
\endgroup
\end{document}